\documentclass{aa} 
\usepackage{graphicx}
\usepackage{txfonts}
\usepackage{natbib}
\usepackage{bm}

\begin{document}

\title{Magnetic field properties of the SNR HB~9 }


\author{
L\sc{i} X\sc{iao},\inst{1,2,3}
    \and M\sc{ing} Z\sc{hu},\inst{1,2,3}
    \and X\sc{iao}-H\sc{ui} S\sc{un},\inst{4}
    \and W\sc{olfgang} R\sc{eich},\inst{5}
    \and P\sc{atricia} R\sc{eich},\inst{5}
    \and P\sc{eng} J\sc{iang},\inst{1,2,3}
    \and C\sc{hun} S\sc{un}\inst{1,2,3}       
}

\offprints{L. Xiao}

\institute{National Astronomical Observatories, Chinese Academy of
            Sciences, Jia-20, Datun Road, 100012 Beijing, China\\
            \email{xl,mz,pjiang,sunchun@nao.cas.cn}
            \and Key Laboratory of FAST, NAOC, Chinese Academy of Science, 100012 Beijing, China
            \and Guizhou Radio Astronomical Observatory, Guizhou University, 550000 Guiyang, China 
            \and School of Physics and Astronomy, Yunnan University, 650091 Kunming, China \\
            \email{xhsun@ynu.edu.cn}
            \and Max-Planck-Institut f\"{u}r Radioastronomie, Auf dem H\"ugel 69, 53121 Bonn, Germany \\
            \email{wreich,preich@mpifr-bonn.mpg.de}  }

\date{Received / Accepted}

\abstract
{}
{We aim to study the polarization and magnetic field properties of the SNR HB~9 using new 21-cm continuum cube data from the Five-hundred-meter Aperture Spherical radio telescope (FAST).
 }
{We computed the Faraday depth at 21~cm, and re-analyzed the rotation measures (RMs) of HB~9 using in addition Effelsberg 2695-MHz
and Urumqi 4800-MHz polarization data. FAST total-intensity images of two subbands are decomposed into components of multiple 
angular scales to check spectral-index variation via temperature versus temperature plots (TT-plots). 
 }
{The filamentary emission has a spectral index ($S\sim\nu^{\alpha}$) of $\alpha=-$0.52, 
corresponding to freshly accelerated relativistic electrons. The diffuse emission has a steeper spectrum of $\alpha=-$0.63, corresponding to confined electrons that are no longer accelerated. 
The FAST detected 1385-MHz polarized emission might come from a thin layer in the outer envelope of the shells, with a Faraday depth of
$4-28$~rad~m$^{-2}$ from the Faraday rotation synthesis result. The RMs derived from the Effelsberg 2695-MHz and Urumqi 4800-MHz polarization data show about 70~rad~m$^{-2}$ in the eastern and northern shell, and 124~rad~m$^{-2}$ in the inner and southern patches. 
The regular magnetic field is about 5$-$8~$\mu$G over the remnant. The northern shell shows depolarization at 2695~MHz 
relative to the 4800-MHz polarization data, indicating an additional random magnetic field of 12~$\mu$G on the scale of 0.6~pc. 
The shock wave might have entered the dense gas environment in the northern-shell region and has driven turbulence to cause depolarization at 2695~MHz.
 }
{}

\keywords{-- Radio continuum: general -- Methods: observational -- ISM: supernova remnants -- ISM: magnetic fields}

\maketitle

\section{Introduction}
Supernova remnants (SNRs) accelerate relativistic particles up to PeV through diffuse shock acceleration (DSA)
mechanism~\citep{be87, Lhaaso21}. 
The accelerated particles escape into the Galactic interstellar medium, making it an important Galactic Cosmic Ray (GCR) source~\citep{b13}.
Radio observations measure the synchrotron emission from accelerated relativistic electrons.
The polarization observations directly detect the magnetic field of the remnant in the sky plane. 
Multi-frequency observations enable to obtain the rotation measures (RMs) through polarization angle $\psi$ variation ($\psi =\psi_{0}+RM~\lambda^{2}$) and derive the magnetic field along the line-of-sight
to construct a three-dimensional configuration of the magnetic field.
This is important for understanding acceleration, diffusion, and evolution of relativistic particles in SNRs.

HB~9 (G160.9+2.6) is a large shell-type SNR with filamentary structures that has been mapped from radio to high-energy $Fermi$ $\gamma$-ray bands~\citep{kff06,lt07,ghr11,a14,syn20}. Its age was estimated to be 4000$-$7000~yrs~\citep{lrg20}.
The distance of the SNR is obtained as 0.54$\pm$0.10~kpc from the optical extinction of background stars~\citep{zjl20}, consistent with the kinematic distance of $0.6\pm0.3$~kpc estimated from the associated HI-shell by~\citet{sey19}.
The pulsar PSR~B0458+46 in the center of HB~9 is excluded from an association with the SNR by a lower limit distance of 2.7~kpc set by HI absorption~\citep{jhh23}, leaving the outside magnetar SGR 0501$+$4516~\citep{gwk10} possibly related.
It is classified as mixed-morphology (MM) SNR with thermal X-ray emission in the center~\citep{la95}.
The $\gamma$-ray spectrum is explained by emission from inverse Compton (IC) scattering of electrons with a simple power-law energy spectrum with an index of 2 and maximum electron energy of 500~GeV~\citep{a14}. 
The magnetic field has been estimated to be 3 to 8$\mu$~G from the spectral energy distribution (SED) fitting~\citep{zxl19, sby22}. 
Accelerated protons and electrons have escaped from the SNR, and interacted with two nearby molecular clouds to produce significant $\gamma$-ray emission~\citep{oi22}.

The radio spectral index ($S\sim\nu^{\alpha}$) varies from $\alpha =-0.57$ to $-0.64$ for frequencies from 408 to 4800~MHz~\citep{rzf03,kff06,ghr11},
with steeper spectra in the interior regions than the filaments of the rim~\citep{lt07}. 
The polarized emission of HB~9 has been previously observed at 21~cm, 11~cm, and 6~cm with the 100-m Effelsberg telescope~\citep{fr04}
to find the intrinsic magnetic field configuration within the SNR. New Urumqi 6-cm polarization observation has improved the signal-to-noise ratio and revealed that the magnetic field mainly runs along the shell~\citep{ghr11}. 

In this paper, we present a new polarization observation of HB~9 at 21~cm with the Five-hundred-meter Aperture Spherical Radio 
Telescope (FAST). FAST has the advantage of a large aperture with high resolution and high sensitivity~\citep{jth20}
and is equipped with a 19-beam wide-band multi-channel polarization receiver from 1050$-$1450~MHz to determine the rotation measures
directly. It enables us to check whether the 21-cm polarized emission comes from the same distance as at shorter wavelengths.
In combination with polarization observations at 11~cm and 6~cm, we analyze the polarization properties, and determine the
rotation measures to estimate the strength of the magnetic field within the remnant.

The paper is organized as follows. The observation and data reduction are described in Sect.~2. 
The decomposition of filaments and diffuse emission and spectral-index analysis is shown in Sect.~3.1. 
The polarization properties at 1385, 2695, and 4800~MHz are described in Sect.~3.2.
The rotation-measure and depolarization analysis are presented in Sect.~3.3 and 3.4. 
In Sect.~4, we discuss the random magnetic field in the northern shell and depolarization associated with a foreground
molecular cloud. Conclusions are summarized in Sect.~5.

\section{Data}
\subsection{FAST 21-cm observations and data reduction}
The FAST continuum mapping of HB~9 was observed in October 2023 in the multibeam 'on-the-fly' mode using the 19-beam L-band receiver. 
It was scanned along the R.A. and Dec. direction centering at $\alpha_{2000}=5^{\rm{h}}01^{\rm{m}}0.0^{\rm{s}}$, $\delta_{2000}$=46$\degr$39$\arcmin$0.0$\arcsec$, with a size of $5.0\degr\times 3.6\degr$. 3C138 was observed as calibrator. 
The spectral backend with 65536 channels covering the 1000$-$1500~MHz band recorded the four polarization outputs, including
left-hand and right-hand total intensity $I_{1}$ and $I_{2}$, and Stokes $U$ and $V$. The Stokes parameters are then calculated as
$I=I_{1}+I_{2}, Q=I_{1}-I_{2}, U$ and $V$. Each channel has a bandwidth of 7.6~kHz.
A noise signal with an amplitude of 10~K was injected every two seconds to calibrate the intensity in the unit of antenna temperature
$(T_{\rm a}$ in Kelvin). The polarization leakage caused by the mismatch of the amplitudes and phases between the gains of the two
linear feeds have been corrected using the injected reference signal.
Radio-frequency-interference (RFI) zone caused by communication satellites and navigation satellites 
(1160-1280~MHz) were flagged, which separates the data into two sub-bands: subband1 centered at 1085~MHz and subband2 at 1385~MHz. 

The data reduction can be summarized as follows: at first, we made a baseline correction to subtract the large-scale emission by linearly
fitting of two ends of each scan. Then, the data are converted to the main-beam brightness temperature ($T_{\rm b}$ in Kelvin)
dividing by the main beam efficiency ($\eta_{b}$) calculated from the calibrator~\citep{smg21}. 
At last, the data are weaved along the R.A. and Dec. direction to destripe the scanning effects in the Fourier domain~\citep{eg88}. 
Each channel map was smoothed to a common angular resolution of 4$\arcmin$. Then they were binned every 20 channels to improve the
noise level.

We averaged the data over the full FAST band by taking medians of all the frequency channels to form the combined Stokes $I$ map. 
For $Q$ and $U$ maps, we only used the subband2 1385~MHz data, which has a better signal-to-noise ratio. 
The polarized intensity $PI$ corrected from the positive noise bias was calculated as $PI = \sqrt{Q^{2}+U^{2}-(1.2\sigma)^{2}}$~\citep{wk74},
where the rms noise $\sigma$ measured from the average combined $Q$, $U$ maps are both about 4~mK. 
The polarization angle $\psi$ is derived as $\psi = \frac{1}{2}\arctan\frac{U}{Q}$. 

\begin{figure}[!hbt]
\begin{center}
\includegraphics[angle=0,width=0.52\textwidth]{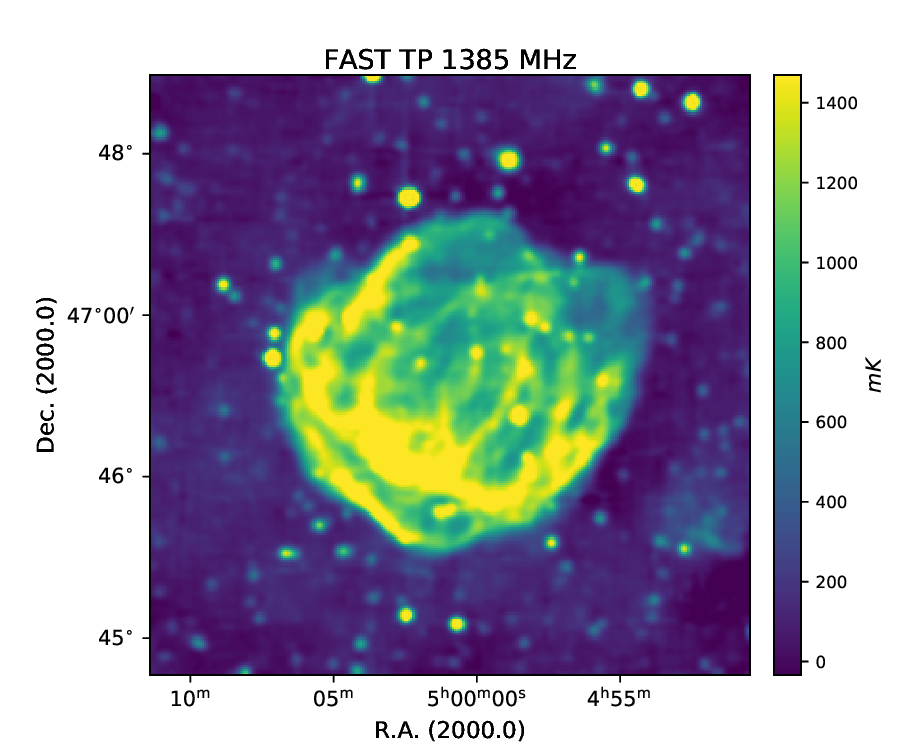}
\includegraphics[angle=0,width=0.52\textwidth]{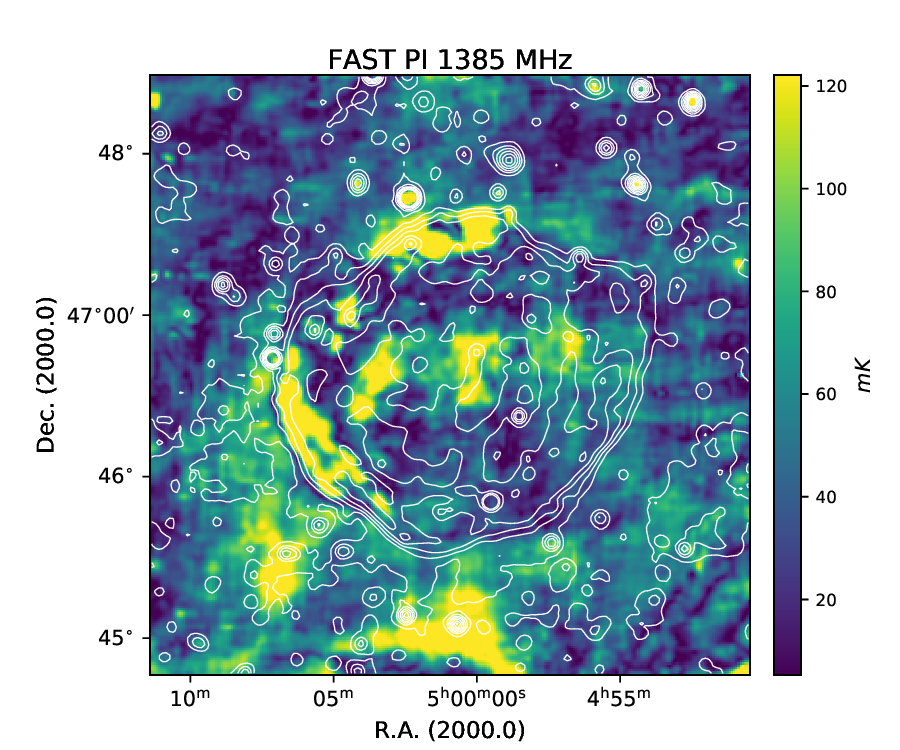}
\caption{$Top~panel$: The FAST total-intensity map of HB~9 of subband2 at a central frequency of 1385~MHz. 
The angular resolution is 4$\arcmin$. $Bottom~panel$: The FAST subband2 polarized-intensity map of HB~9 at 1385~MHz.  
Contours show the total intensities of HB~9 at 150, 400, 700, 1100~mK and further increase in steps of 500~mK. 
}
\label{fast_pi}
\end{center}
\end{figure}

\begin{figure*}[!hbt]
\begin{center}
\includegraphics[angle=0,width=1.\textwidth]{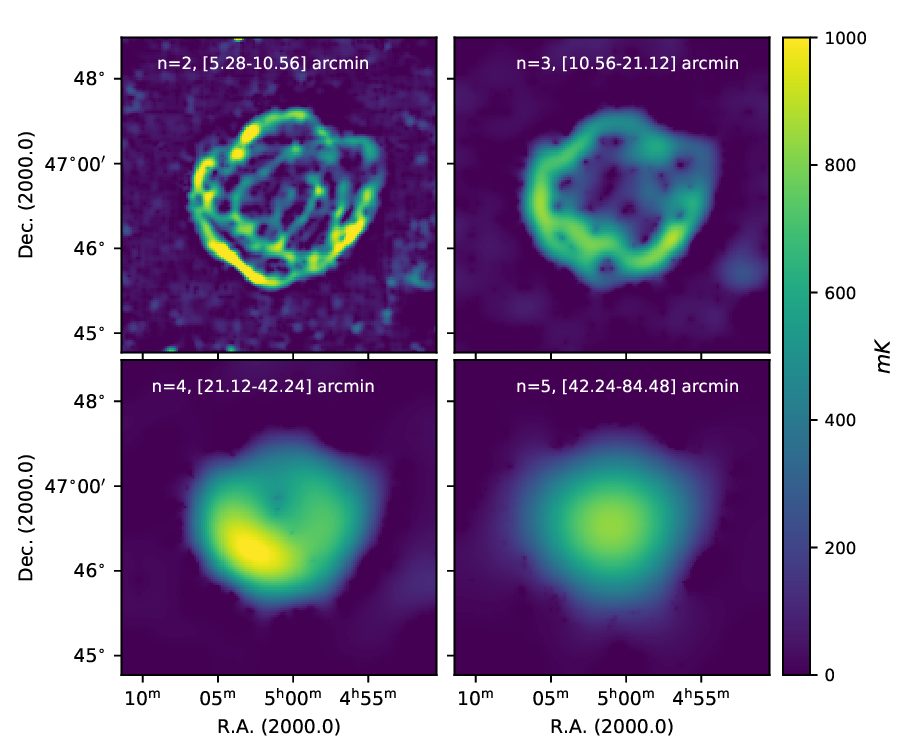}
\caption{Component images from the constrained diffusion decomposition (CDD) of HB~9 at FAST subband at 1085~MHz. 
The color scale extends from 0 to 1.0~K~$T_{\rm b}$.
The component number $n$ and the corresponding range of the angular scales are marked in each panel. For component n$<$2, the angular scale is less than the beam size, and the emission is weak, hence it is not shown.
}
\label{Dcomp}
\end{center}
\end{figure*}

The FAST averaged subband2 total-intensity and polarization maps of HB~9 at 1385~MHz are presented in Fig.~\ref{fast_pi}.
The shell and the filamentary structures are clearly identified. 
Polarization emission has been detected in the eastern filamentary shell, the diffuse end of the northern shell, and the diffuse end of the inner filamentary shell at a similar level of 160~mK $T_{\rm b}$. The FAST 1385-MHz polarization-emission morphology is similar compared with the Dominion Radio Astrophysical Observatory (DRAO) 1420-MHz
polarization-intensity map~\citep{kff06}, except for no detection of weak polarization emission in the southern filamentary shell region.
This is reasonable as the DRAO synthesis map has a high angular resolution of $\sim 1\arcmin$, which
suffers less beam depolarization.

\subsection{Effelsberg 2695-MHz and Urumqi 4800-MHz maps}
The archival total-intensity and polarization maps of HB~9 at 2695~MHz were observed with the Effelsberg 100-m telescope
during test observations for the Effelsberg Galactic plane survey~\citep{rfh84}, where observation methods and data reduction 
were described. The angular resolution is 4.3$\arcmin$. The 2695-MHz map was already used for a RM study of HB~9 using Effelsberg
1420-MHz polarization data as second frequency~\citep{fr04}.
The data at 4800~MHz were from the `Sino-German' $\lambda$6~cm polarization survey of the Galactic plane~\citep{grh10}.  
The angular resolution is about 9.75$\arcmin$. It is publicly available in the ``Survey Sampler" 
of the Max-Plank-Institut f\"ur Radioastronomie \footnote{http://www.mpifr-bonn.mpg.de/survey.html}.
Considering a shock velocity of 300~km~s$^{-1}$ in the Sedov phase deduced from the maximum velocity observed in
H$\alpha$ line~\citep{llm24}, 
HB~9 will expand less than 5$\arcsec$ in 40~yrs, which is relatively static at the angular resolution of 4.3$\arcmin$.
It is not a restriction when compared with the recent FAST map.

\begin{figure*}[!hbt]
\begin{center}
\includegraphics[angle=0,width=0.3\textwidth]{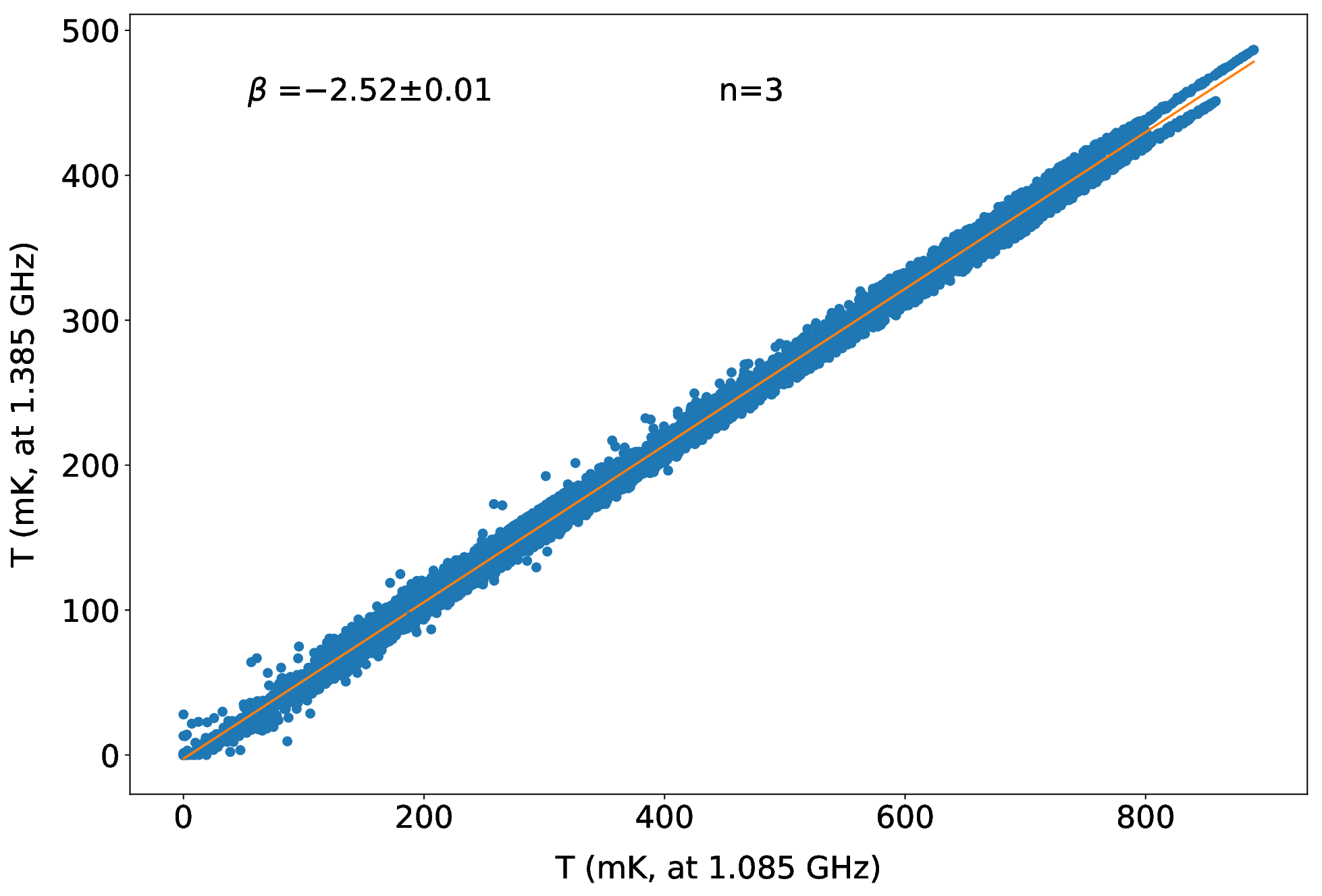} 
\includegraphics[angle=0,width=0.3\textwidth]{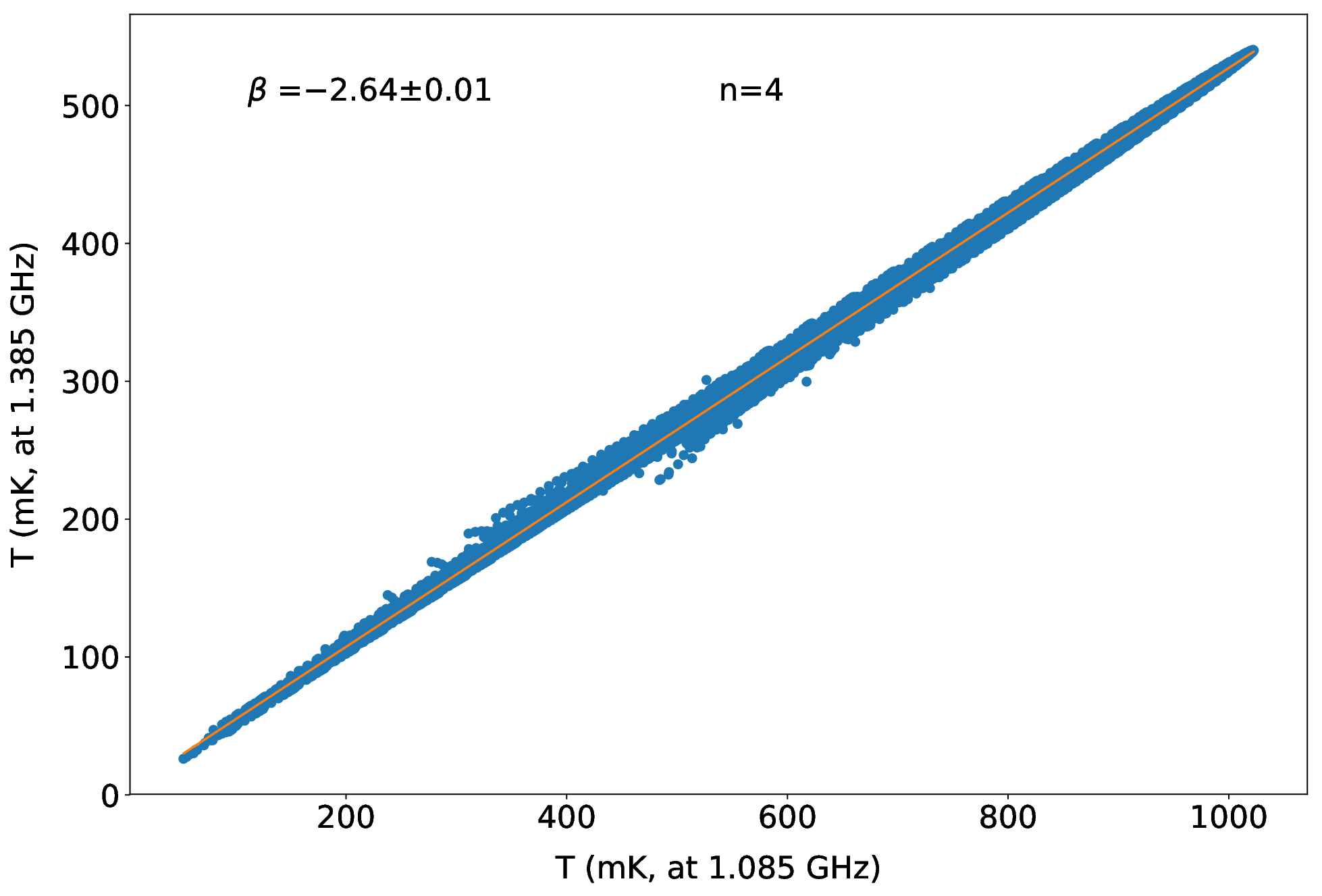}
\includegraphics[angle=0,width=0.3\textwidth]{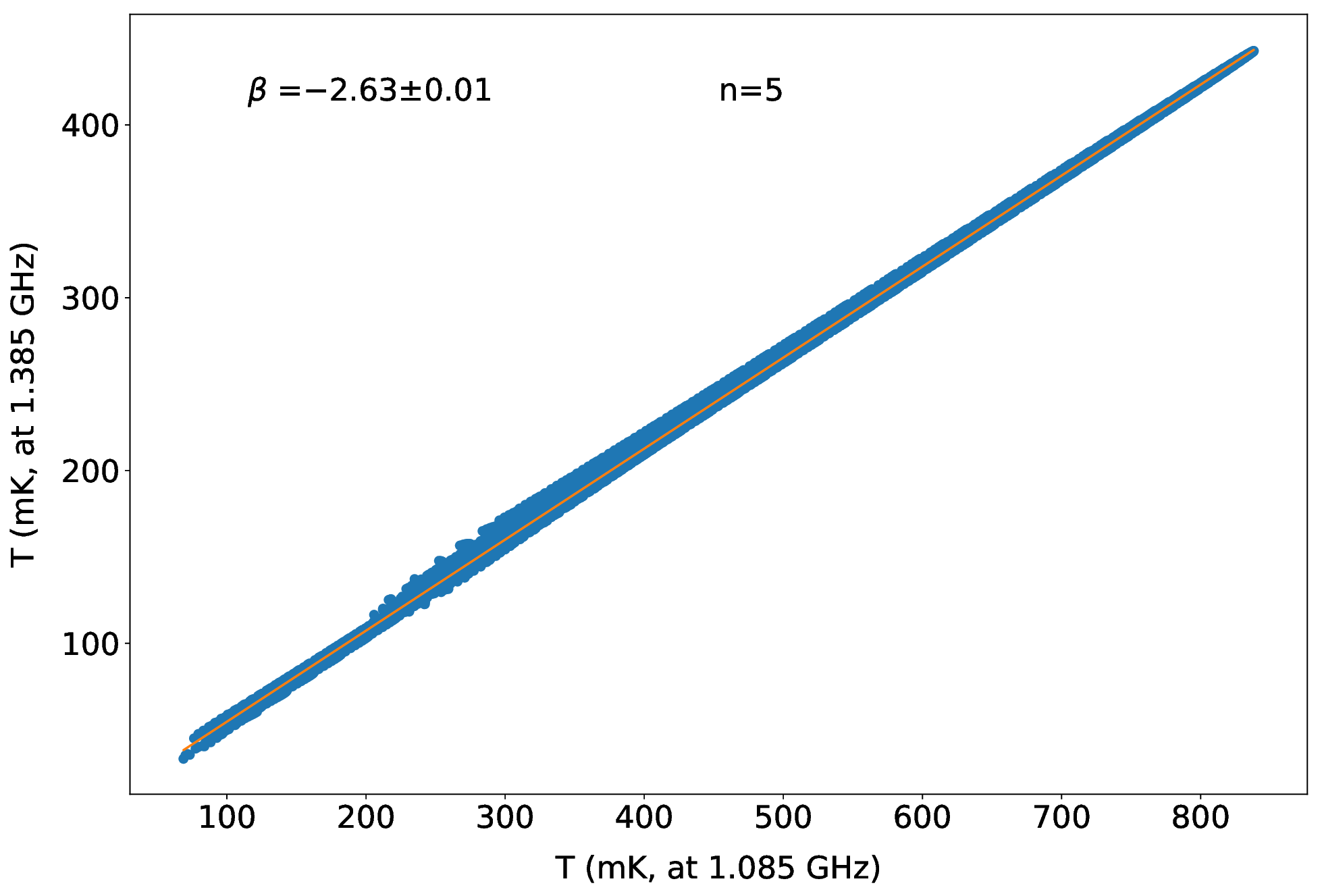}
\caption{Temperature versus temperature plots (TT-plots) between FAST 1085-MHz and 1385-MHz subbands for total-intensity components at $n=3$ to $n=5$. $\beta$ is derived from the data's slope $(\nu_{y}/\nu_{x})^{\beta}$, and the error is estimated by fitting the data twice, taking the two wavelengths as the independent variable individually.
}
\label{index}
\end{center}
\end{figure*}

\subsection{CO data for comparison}
We used the $^{12}$CO (J=1$-$0) data from~\citet{zsy23} to check the molecular cloud environment around HB~9, which were obtained from
the Milky Way Image Scroll Painting (MWISP)$-$CO line survey project using the Purple Mountain Observatory Delingha 13.7~m
millimetre wavelength telescope. The angular resolution is $\sim$51$\arcsec$, and the data were meshed with a grid spacing of 30$\arcsec$.
The typical rms noise level is about 0.5~K at the spectral resolution of 0.16 km~s$^{-1}$. 
We integrated the $^{12}$CO (J=1$-$0) emission in the velocity ranges of [$-$10.0, +2.0]~km~s$^{-1}$ and [$-$11.5, $-$10.0]~km~s$^{-1}$,
respectively, and compared it with the FAST 1385-MHz data in Sect.~4.1.

\section{Results}
\subsection{Spectral index in the filament and diffuse emission}
We applied the constrained diffusion decomposition (CDD) method~\citep{l22} to decompose the total-intensity image of 
Fig.~\ref{fast_pi} into components of multiple angular scales.
The CDD method improves the image smoothing quality by solving a modified, nonlinear version of the diffusion equation,
and does not produce artifacts around regions with sharp transitions.
The component images are shown in Fig.~\ref{Dcomp}. 
For component n, the corresponding scale of the structures is larger than 2$^{n}\Delta$ but smaller than $2^{n+1}\Delta$, 
where $\Delta=1.32\arcmin$ is the grid size of the map.
The bright filamentary shell structures predominantly are visible in components n=2 and n=3. For component n$<$2, the angular scale is less than the beam size, and the emission is very weak. Hence, these components are not shown.

We made temperature versus temperature plots (TT-plots) to determine the brightness temperature spectral index $\beta$, 
defined as $T_{\nu}\sim \nu^{\beta}$. $\beta$ is derived from the data's slope $(\nu_{y}/\nu_{x})^{\beta}$, and the error
is estimated by fitting the data twice, taking the two wavelengths as the independent variable individually.
This method is less affected by the uncertainty of the baselevels at these two frequencies.
The flux-density spectral index $\alpha$, defined as $S_{\nu}\sim \nu^{\alpha}$
with $S_{\nu}$ being the flux density, can then be obtained as $\alpha=\beta +2$. We used the two averaged subband data at 1085~MHz 
and 1385~MHz to derive TT-plots for different components of $n>2$ (to avoid beam averaging). 
The point-like sources have been subtracted from the map and the results are shown in Fig.~\ref{index}. 
The filamentary and diffuse emission of the SNR show different spectral indices.

The spectral indices of the filamentary emission of n=3 components are around $\alpha\sim-0.52$, which is the standard DSA 
theory~\citep{bo78} expected for freshly accelerated relativistic electrons by the shock in the filaments.
While the spectral indices of the diffuse emission of components n=4 and n=5 with angular scales larger than 21$\arcmin$ are 
$\alpha\sim-0.63$. This trend is similar to the result found by~\citet{lt07}. The energetic accelerated energetic particles
have escaped away from the remnant. The spectrum of confined particles in the diffuse emission, which are no longer accelerated,
becomes steeper~\citep{sby22}.

\begin{figure*}[!hbt]
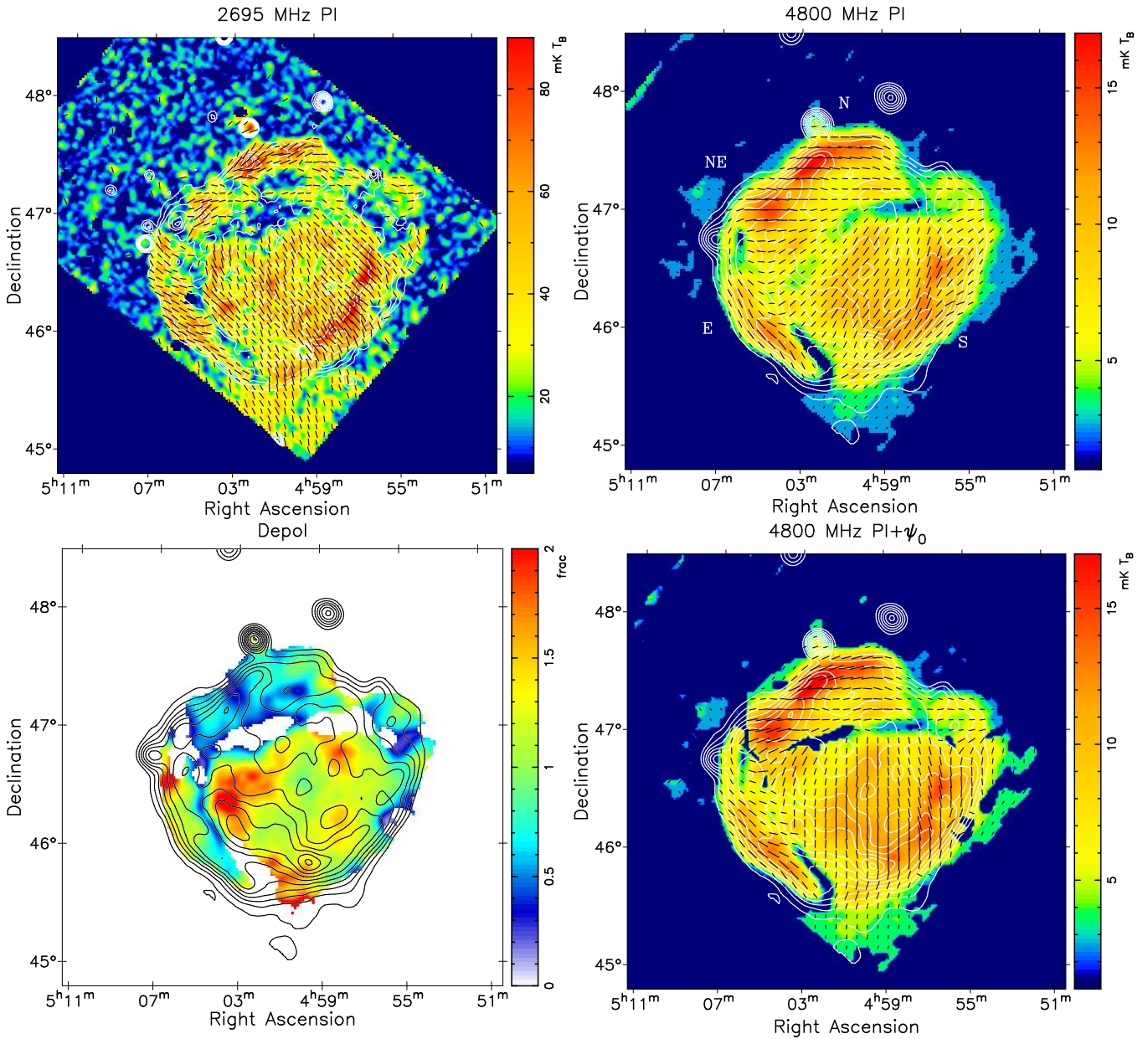

\begin{center}
\includegraphics[angle=-90,width=0.48\textwidth]{G160_11cmpi.radec-eff-TAN.ps}
\includegraphics[angle=-90,width=0.48\textwidth]{G160_6cmpi.radec-TAN.ps}
\includegraphics[angle=-90,width=0.48\textwidth]{G160_dp.radec-TAN.ps}
\includegraphics[angle=-90,width=0.48\textwidth]{G160_6cmchi0.radec-TAN.ps}
\caption{$Upper~panels$: The Effelsberg 2695-MHz and Urumqi 4800-MHz polarized-intensity map of HB~9, overlaid with bars indicating
B vectors (polarization angle $+ 90\degr$ in case of negligible Faraday rotation). The contours in the left panel show the 
total intensity at 2695~MHz, starting at 100~mK in steps of 60~mK. The contours in the right panel show the total intensity at 
4800~MHz (right), starting at 15~mK in steps of 8~mK. $Lower~panels$: The depolarization factor map at 2695~MHz of HB~9 ($left$), 
derived from the polarization percentage maps at 2695 and 4800~MHz ($PC_{2695}/PC_{4800}$). Values $\ge$ 1 mean no depolarization.
A value of 0 means complete depolarization. The white areas are low-polarized intensity regions excluded from the depolarization calculation.
The $right$ panel shows the Urumqi 4800-MHz polarized intensity map of HB~9 with overlaid bars showing the intrinsic orientation of the transverse magnetic field.
}
\label{pi_6_11}
\end{center}
\end{figure*}

\subsection{Polarization maps at 1385, 2695, and 4800~MHz}
As shown in the FAST 1385-MHz map in Fig.~\ref{fast_pi}, polarization emission from the bright filaments in each shell
is depolarized, which indicates that the observed polarization emission at 1385~MHz might come from a thin layer in the outer envelope 
of the shell, and can not be used for the RM calculation with 2695-MHz and 4800-MHz data. 
In the diffuse end of the northern shell, there is also a depolarized zone (discussed in detail in Sect. 4.2).

The Effelsberg polarized-intensity map of HB~9 at 2695~MHz and the Urumqi polarized-intensity map at 4800~MHz are presented in Fig.~\ref{pi_6_11}. The polarized intensity at 2695~MHz corresponds well to that at 4800~MHz with strong polarized emission confined to
the outer and inner filamentary shock shells. They are distinct at different sections (labeled E, NE, N, and S). 
The polarization angle at 4800~MHz mainly runs parallel along the shock shell indicating the evolved properties of HB~9. 
The polarization percentage $PC$ at 4800~MHz in the northeastern and northern shell region is about 20\%$-$40\%, $\sim$22\% in the eastern region, $\sim$24\% in the southern region, and a lower value in the central region. 

\begin{figure*}[!hbt]
\begin{center}
\includegraphics[angle=0,width=0.49\textwidth]{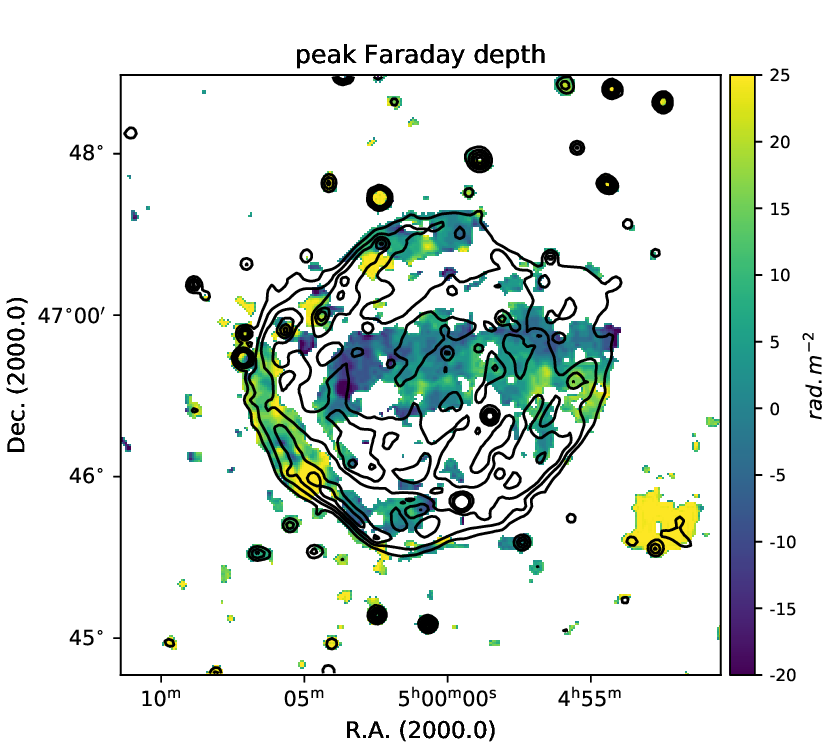}
\includegraphics[angle=0,width=0.49\textwidth]{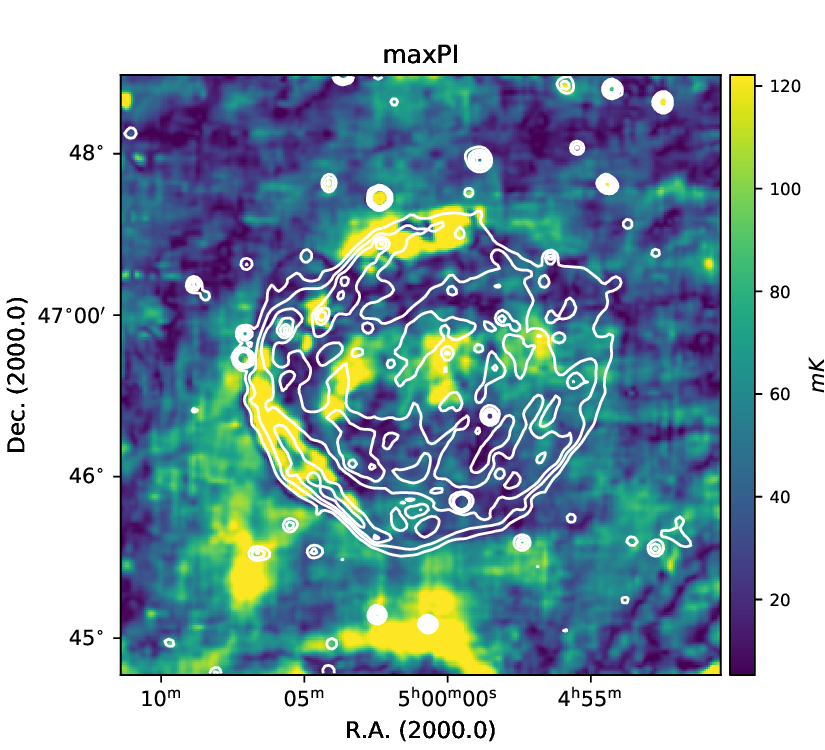}

\caption{The RM ($left~panel$) and maximum polarized intensity ($right~panel$) of HB~9 at 1240~MHz derived from the RM synthesis method. The contours show the subband2 total intensity at 1385~MHz,
starting at 500~mK in steps of 400~mK. 
}
\label{peakRM}
\end{center}
\end{figure*}

Assuming that the depolarization is the least at 4800~MHz and using $PI_{4800}$ as reference, we can check the depolarization
properties at 2695~MHz by calculating the relative polarization-percentage map interpreted as the depolarization factor of
$DP_{2695}=PC_{2695}/PC_{4800}$. If there is no depolarization, $DP_{2695}$ would be around 1.
The result is depicted in the lower left panel of Fig.~\ref{pi_6_11}. The depolarization factor is around 1 for all areas except for the
northeastern and northern shells, where $DP_{2695}$ varies from about 0.6 to about 0.7, indicating depolarization towards this region.

\subsection{RM determination}
\subsubsection{RM-Synthesis result of 1385-MHz data}
We applied the RM-synthesis method to the FAST frequency cubes of $Q$ and $U$ of HB~9 to reconstruct the Faraday-depth spectrum~\citep{bd05}.
The method is based on that the observed polarized intensity $P(\lambda^{2})$ and the Faraday dispersion function $F(\phi)$ are
Fourier transform pairs~\citep{b66} 

\begin{equation}\label{RMsyn}
P(\lambda^{2})= Q(\lambda^{2})+iU(\lambda^{2})=\int F(\phi)e^{2i\phi\lambda^{2}}d\phi,  
\end{equation}
where $\lambda$ is the wavelength and the $\phi$ is the Faraday depth defined 
as $\phi=0.812\int n_{e}B_{||}dr$, with the thermal electron density $n_{e}$ in cm$^{-3}$, the magnetic field along the line-of-sight $B_{||}$ in $\mu$G, 
and the differential of the path length from the source to the observer $dr$ in pc.

The algorithm is applied for each pixel using the RM-Tools package~\citep{pvw20} adopting a Faraday-depth step interval of 4~rad~m$^{-2}$.
For the subband2 frequency range and sampling, the width of the RM spread function is about 280~rad~m$^{-2}$.
It returns the peak Faraday depth $\phi_{peak}$, which is equivalent to the rotation measure in rad~m$^{-2}$, 
and the maximum polarized intensity $F(\phi_{peak})$ (maxPI) at the peak Faraday depth.   
 
We display the maxPI map and peak Faraday depth $\phi_{peak}$ map towards HB~9 in Fig.~\ref{peakRM}. 
The polarized intensities in the maxPI map are nearly identical to the value directly obtained from the
averaged $Q$ and $U$ in the lower panel of Fig.~\ref{fast_pi}, indicating the bandwidth depolarization is negligible during averaging.
Most of the Faraday spectra show a single peak with only one emitting component, as the example spectra presented in~\citet{smg21}.
The peak RM in the eastern, northern shell, and inner region varies from 4 to 28~rad~m$^{-2}$. 
The RM contribution from the foreground interstellar medium can be neglected. Towards the direction of HB~9,
the large-scale magnetic field component parallel to the line-of-sight $B_{||}$ is close to 0 as it runs along the spiral arm.
Assuming an electron density $n_{e}\sim 0.02$~cm$^{-3}$, and a magnetic field strength of $2~\mu$G as typical values for the local interstellar medium, the foreground RM would be less than 2~rad~m$^{-2}$ at a distance of 540~pc.

\begin{figure}[!hbt]
\begin{center}
\includegraphics[angle=-90,width=0.48\textwidth]{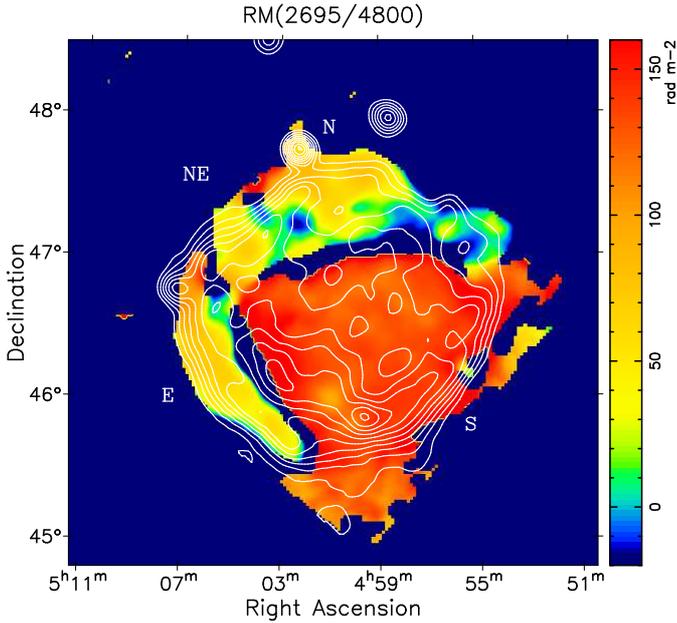}
\caption{The RM map of HB~9 calculated from the 2695-MHz and 4800-MHz maps. The contours show the total intensity at 4800~MHz, 
starting at 15~mK in steps of 8~mK. The eastern and northern shells have RMs of 80~rad~m$^{-2}$, while the inner and southern shells have RMs of 124~rad~m$^{-2}$.
}
\label{rm}
\end{center}
\end{figure}

\subsubsection{Rotation measures between 2695~MHz and 4800~MHz}
We used the conventional method to calculate the RM from the polarization angle variation caused by Faraday rotation at different 
wavelengths $\lambda$. Observed RMs correspond to the Faraday depth $\phi$. Positive/negative RMs mean that the magnetic field
is pointing towards/away from the observer.

The rotation measure calculated from data at 4800~MHz and 2695~MHz is 
RM$=\frac{\psi_{2}-\psi_{1}}{\lambda_{2}^{2}-\lambda_{1}^{2}}$, where the subscripts 1 and 2 stand for quantities at these 
two wavelengths, respectively. There is a $n\pi$ ambiguity of polarization angles, resulting in a RM ambiguity of 
$n\pi/(\lambda_{2}^{2}-\lambda_{1}^{2})\sim 370$~rad~m$^{-2}$, $n=0, \pm 1, \pm 2, ...$.
We smoothed the 2695-MHz $U$ and $Q$ maps to the same angular resolution of 9.75$\arcmin$ at 4800~MHz and calculated the $PI$ map.
The RMs were obtained for each pixel from all maps, except those with polarized intensity less than the 
5$\sigma$ level, which is about 50~mK at 2695~MHz and 1.5~mK at 4800~MHz. 
The minimal RMs map of HB~9 is presented in Fig.~\ref{rm}. The RM in the eastern shell is about 70~rad~m$^{-2}$ (E), and the northern shells
have similar RMs of about 70$-$80~rad~m$^{-2}$ (NW), while the northeastern shell has a value of 60$-$70~rad~m$^{-2}$.
The inner southern shell (S) has an RM of 124~rad~m$^{-2}$. The intrinsic polarization angles $\psi_{0}$ determined based on RMs
are shown in lower right panel of Fig.~\ref{pi_6_11}. The intrinsic perpendicular magnetic field aligns well along the shell 
in the E, NE and N region.

According to the simulation of a spherical SNR expanding into an ambient medium of constant density and homogeneous
magnetic field~\citep{kb09}, the RM values on both shells almost have the same value.
For the inner southern shell of HB~9, the RM (n$=-1$) of $-$240~rad~m$^{-2}$ could be excluded.
Larger RM values (n$=1$) are also impossible, considering the physical conditions in the shell.

\begin{figure}[!hbt]
\begin{center}
\includegraphics[angle=0,width=0.49\textwidth]{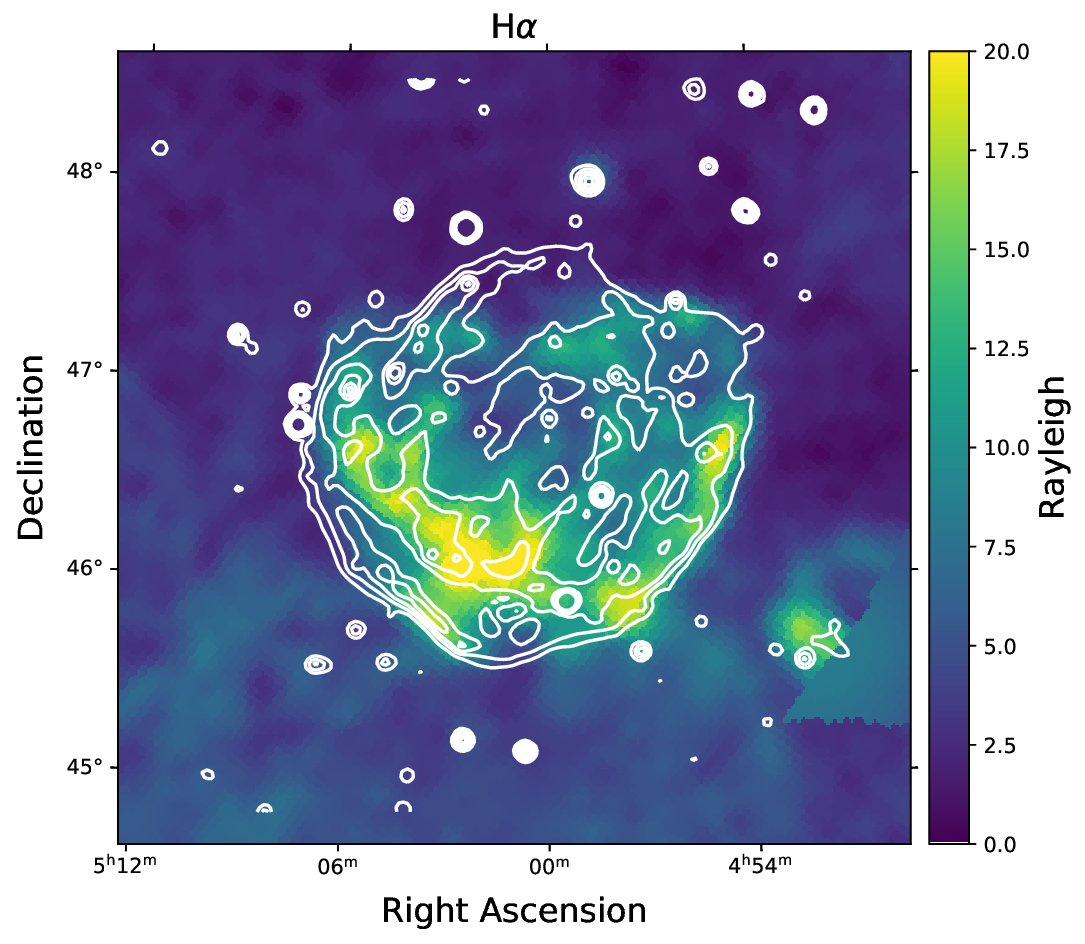}
\caption{H$_{\alpha}$ map of HB~9~\citep{f03}. The overlaid contours show the FAST subband2 total intensity at 1385~MHz, starting at 500 mK in steps of 400 mK.
}
\label{Halpha}
\end{center}
\end{figure}

\subsection{Depolarization analysis}
The northeastern and northern shells show a depolarization factor between 2695~MHz and 4800~MHz of 0.6$-$0.7, while the other regions
have a value of 1 (bottom left image in Fig.~\ref{pi_6_11}). We considered the possible depolarization mechanism there. According to ~\citet{sbs98}, two main depolarization mechanisms 
may cause depolarization for the synchrotron emitting region.
One is depth depolarization, caused by thermal gas co-existing with the synchrotron emitting region.
The other is beam depolarization caused by RM fluctuations ($\delta_{RM}$) within the beamsize of the telescope in front of the synchrotron-emission medium.
As the other shells with similar or larger observed rotation measures show no depolarization at 2695~MHz, 
we considered beam depolarization as the main depolarization mechanism in the northeastern and northern shells.

The depolarization factor of beam depolarization can be written as $DP_{\nu}=exp[-2\delta^2_{RM}(\lambda^{4}-\lambda^{4}_{0})]$~\citep{sbs98}. 
The values of $DP_{2695}$ yield RM fluctuations from $\sim 36$~rad~m$^{-2}$ to $\sim$ 43~rad~m$^{-2}$.
The RM fluctuation at these levels will cause virtually complete depolarization at 1385~MHz. 
We noticed that the northeastern and northern shells were in contact with a molecular cloud (Sect.~4.1). 
It is possible that shock waves have entered the dense gas environment in these regions, and have driven turbulence and caused 
the RM dispersion to result in depolarization.

\begin{figure*}[!hbt]
\begin{center}
\includegraphics[angle=0,width=0.46\textwidth]{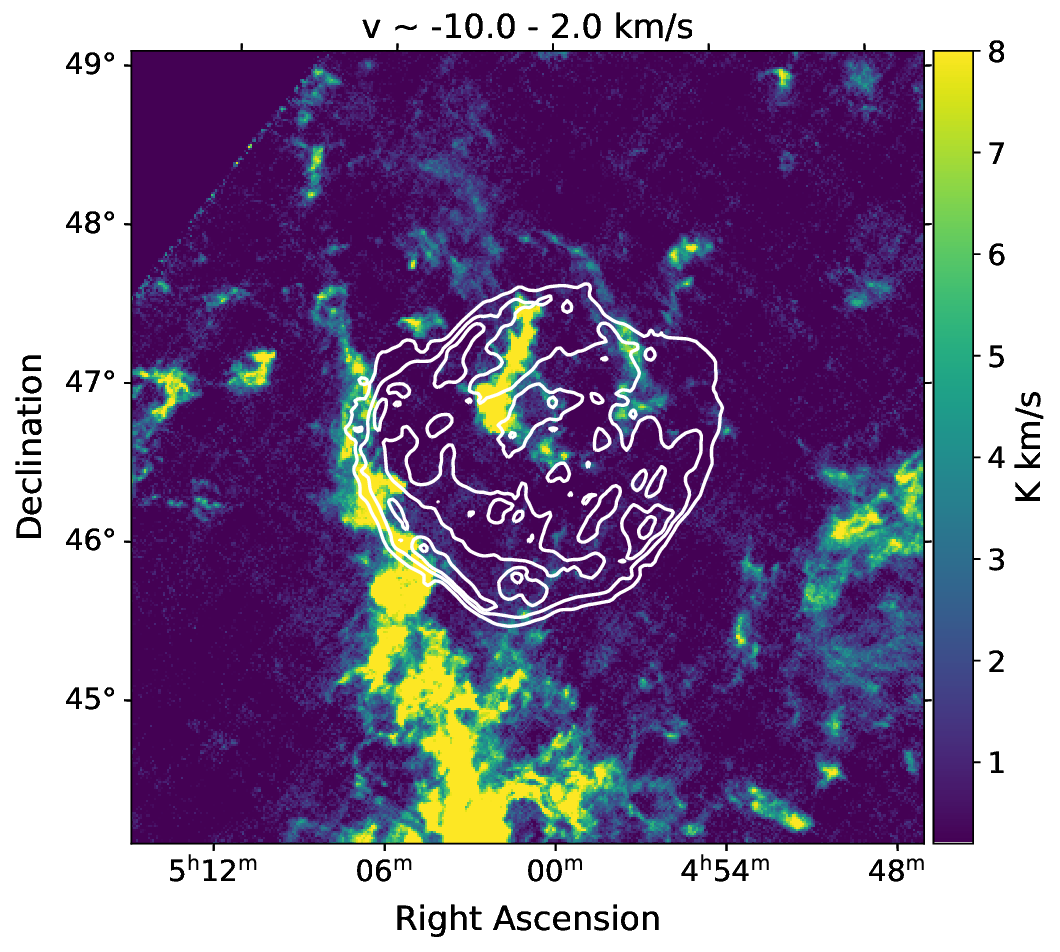}
\includegraphics[angle=0,width=0.46\textwidth]{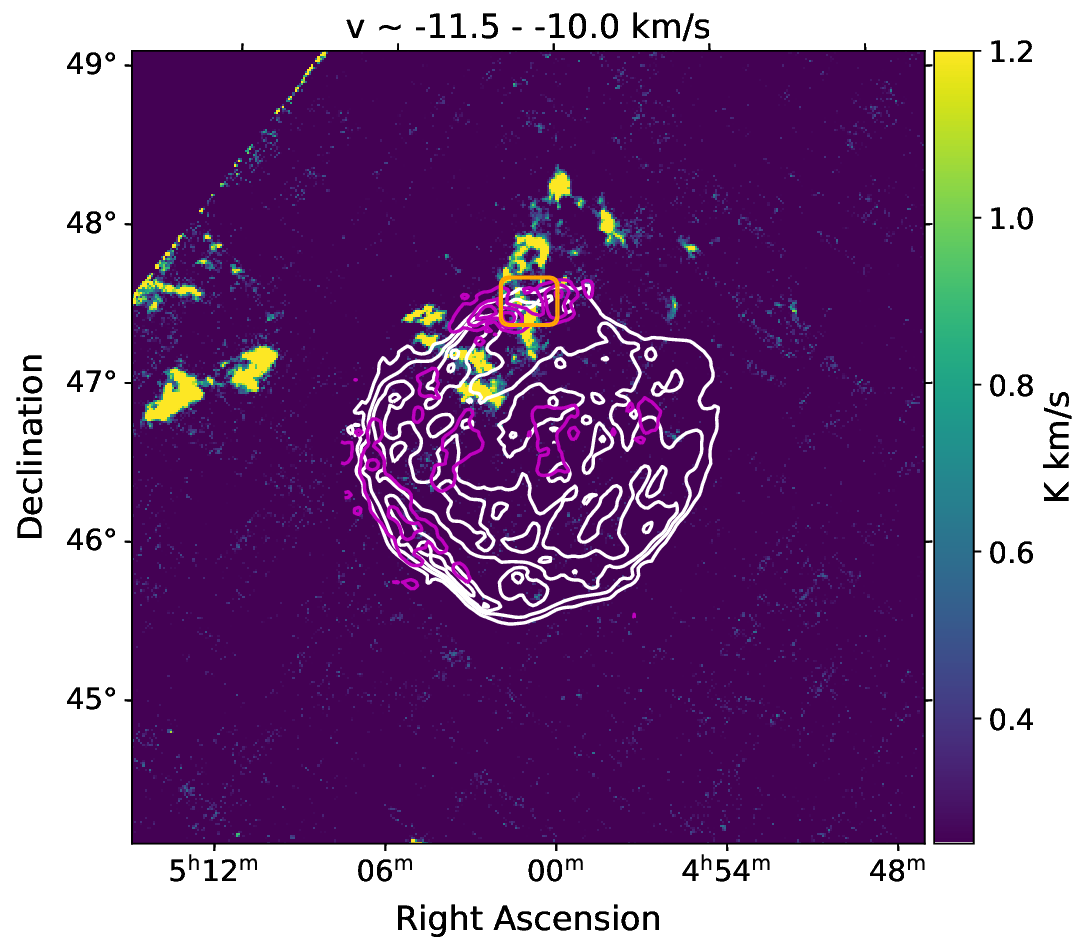}
\caption{CO map in the velocity range of $-$10.0 to $+$2.0~km~s$^{-1}$ ($left$) and $-$11.5 to $-$10.0~km~s$^{-1}$ ($right$) of HB~9. 
White contours show the subband2 total intensity of HB~9 at 1385~MHz with point sources subtracted,
starting at 400~mK in steps of 500~mK.
Magenta contours show the subband2 polarized intensity at 1385~MHz, starting at 100~mK in steps of 50~mK. 
The orange box marks the polarization void region at R.A.$\sim 5^{\rm{h}}01^{\rm{m}}$, Dec.$\sim 47\degr 33\arcmin$.
}
\label{co}
\end{center}
\end{figure*}

\subsection{The regular and random B strength}
From the H$_{\alpha}$ map shown in Fig.~\ref{Halpha}~\citep{f03}, the H$_{\alpha}$ intensity is about 18~Rayleigh towards the 
southeastern shell. Following~\citet{hrt98} and neglecting absorption, the emission
measure ($EM$) can be derived as $EM=2.75T_{4}^{0.9}I_{H_{\alpha}}$. Here $EM$ is in pc~cm$^{-6}$, the electron temperature
$T_{4}$ is in 10$^{4}$~K, and $I_{H_{\alpha}}$ is in Rayleigh. Taking the typical value of $T_{4}=0.8$, we obtained the $EM$ of
40~pc~cm$^{-6}$. The width of the filamentary shell is about 5$\arcmin$, which corresponds to 0.8~pc (adopting the distance
of 540~pc), and the path length along the line-of-sight $l$ is about 7.5~pc.
The thermal electron density $n_{e}$ can be derived according to $EM\sim$$n^2_{e}l$, which is about 2.3~cm$^{-3}$ in the eastern shell. 
Here the volume filling factor for the thermal electrons is assumed to be 1. 
The optical [SII]~6716/6730 line ratio shows $n_{e}$ variations from the inner to outer eastern shell~\citep{llm24}. 

The H$_{\alpha}$ intensities in the northern, inner shell are about 10 and 22~Rayleigh, corresponding to
$EM$ of 22 and 50~pc~cm$^{-6}$ respectively, and the electron density of 1.7 and 2.6~cm$^{-3}$ assuming a similar path length of the shell.
The average RM derived from the data at 2695~MHz and 4800~MHz is about $70$~rad~m$^{-2}$ towards the eastern and northern shell, without foreground contributions. 
The regular magnetic field there can then be estimated as about 5.0 and 6.8~$\mu$G.  
With RM of 124~rad~m$^{-2}$ in the inner shell region, the regular magnetic field there is about 7.6~$\mu$G. 
This estimate largely depends on the geometry path length of the shell, with a relation of $B_{||}=\frac{RM}{0.81n_{e}l}=\frac{RM}{0.81\sqrt{EM~l}} \sim l^{-0.5}$.
~\citet{zxl19} modeled the SED spectrum of HB~9 from radio to $\gamma$-ray bands, and obtained a magnetic field of about 3~$\mu$G 
for the whole SNR. ~\citet{oi22} fitted a magnetic field of 8~$\mu$G from the Fermi-LAT $\gamma$-ray emission spectra around HB~9. 
Our results are more inclined towards the latter, which only considers the inverse Compton process and has fewer model parameters,
indicating that the leptonic process dominates HB~9.

The RM fluctuation towards the northern shell can be interpreted as $\delta RM\sim 0.81 n_{e}b\sqrt{Ll}$~\citep{srh11}, 
where $b$ represents the strength of the random magnetic field in $\mu$G, and $L$ is the correlation scale for magnetic field fluctuations 
in pc. For beam depolarization, $L$ should be much smaller than the beam width of 4$\arcmin$ or 0.6~pc. 
The path length along the line-of-sight $l$ is similar as the eastern shell of 7.5~pc.
The lower limit of $b$ is about 12~$\mu$G, larger than the regular magnetic fields.

\section{Discussion}
As shown by~\citet{sey19}, HB~9 is expanding into dense CO and HI material in the northeastern/eastern and southern sides. 
The ambient gas environment will affect the magnetic field and polarization properties of SNRs.
The enhanced regular magnetic field in the CO and HI structures can make them act as 
Faraday screens to cause depolarization~\citep{wr04,xzs23}. In this section, we will compare the radio map with new CO data 
obtained by the MWISP survey to discuss the random magnetic field in the northern shell of HB~9 and to analyze the 
depolarization void in the northern shell. We present CO maps in the velocity range of $-$10.0 to $+$2.0~km~s$^{-1}$ (left panel) and
$-$11.5 to $-$10.0~km~s$^{-1}$ (right panel) in Fig.~\ref{co}, with the FAST 1385~MHz emission of HB~9 overlaid as contours.

\subsection{The random magnetic field in the northern shell}
From the integrated broad velocity intervals [$-$10.0 $-$ $+$2.0]~km~s$^{-1}$ map of the MWISP CO data in the left panel of Fig.~\ref{co},
the northeastern shell of HB~9 is adjacent to the tail of the eastern CO cloud. Enhanced dense HI gas well follows the northeastern shell
in Fig.~4 of~\citet{sey19}. Although no broadened CO line profile was found, it indicates that the shock wave 
is expanding into a dense gas environment.

When a shock encounters a density enhancement, the shock changes from nonradiative to radiative.
Thermal instabilities can develop in shocks faster than about 120~km~s$^{-1}$~\citep{igf87}.
Any density inhomogeneities in the upstream gas can generate velocity shear and vorticity, 
creating small-scale turbulence, winding up, and amplifying the magnetic field~\citep{gj07,gll12}. 
It will cause a disordered magnetic field and reduce the degree of polarization of radio synchrotron emission.

A similar turbulence has been reported in the eastern and western shell of Cygnus Loop, where the remnant drives 100$-$150~km~s$^{-1}$
radiative shock into high-density clouds~\citep{rsb20}. It leads to a full depolarization at 21~cm, and a random magnetic field 
lower limit of 5~$\mu$G towards the northeastern part and 3~$\mu$G towards the western part~\citep{sgr22}.
For HB~9, the shock possibly has entered the dense gas environment in the northeastern shell and has driven the turbulence magnetic field,
which caused the depolarization at 2695~MHz.

\subsection{Depolarization feature associated with the molecular cloud}
The depolarization feature in the northern filamentary shell at R.A.$\sim 5^{\rm{h}}01^{\rm{m}}$, Dec.$\sim 47\degr 33\arcmin$
coincides with a spherical CO cloud at $-$11.5 to $-$10.0~km~s$^{-1}$ (orange box in the right panel of Fig.~\ref{co}) with a 
width of 10$\arcmin$. Considering a cylinder shape, the physical size of the CO cloud is of the order of 1.6~pc for the distance
of HB~9 of 540~pc. There are also depolarization structures revealed in the same region of the CGPS map~\citep{kff06}. 

It has been reported that the surface of the local molecular clouds act as magneto-ionic Faraday screens to cause depolarization
at 21~cm by rotating background emission adding to foreground emission in a different way than the surroundings~\citep{wr04}.
\citet{ft95} show that the local synchrotron emissivity has a strong linear variation with distance.
Several Faraday-screen modelling works have found local emissivities at 1.4~GHz of 1.7K/kpc~\citep{wr04},
0.94~K/kpc or 0.70~K/kpc for G107.0+9.0, depending on the distance~\citep{rgr21}, and 1.1 K/kpc for G203.7+11.5~\citep{rrs20}.
Adopting a typical emissivity of 1 K/kpc at 1.4 GHz towards local objects,
the foreground emission in front of HB~9 (d$\sim$540~pc) is about 540~mK. It corresponds to polarized emission of about 160~mK for 
a~30\% percentage, which is comparable to the emision towards the northern shock shell.
Thus it is possible that the CO cloud has acted as a Faraday screen along its periphery. 
The background shell emission is rotated and canceled adding to foreground emission.
The enhanced B-field could come from the strong stellar wind of the progenitor star of HB~9, or the formation process of the 
molecular cloud. The enhanced electron density $n_{e}$ comes from the carbon and hydrogen in a thin outer layer in the cloud,
photodissociated by the hydrogen-ionizing photons from the progenitor star and cosmic rays from the remnant. 

We estimated the total electron density $n_{e}$ in the boundaries of the cloud in the following way.
The excess reddening at the position of the CO cloud is $E_{B-V}\sim 0.8$~mag~\citep{sfd98}.
The total neutral-hydrogen density amounts to $n_{(HI+H_{2})}\sim$ 940~cm$^{-3}$ for the size of the cloud of 1.6~pc,
based on the gas-to-dust relation $N_{(HI+H_{2})}/E_{B-V}=5.8\times 10^{21}$~cm$^{-2}$mag$^{-1}$~\citep{bsd78},
where the $N_{(HI+H_{2})}$ is the total neutral-hydrogen column density.
The $C/H$ ratio in the cloud is related to the fractional abundances of molecular hydrogen 
$f(H_{2})=\frac{2N_{H_{2}}}{N_{HI}+2N_{H_{2}}}$~\citep{g00}. For extinctions of $A_{V}\sim 3.1E_{B-V}\sim 2.48$~mag, 
the PDR model of~\citet{htt91} gives $f(H_{2})>10^{-3}$, which corresponds to a $C/H$ ratio of $1.36\times$10$^{-4}$. 
The local electron density from the complete ionization of carbon will contribute an enhancement of about 0.13~cm$^{-3}$.
As the UV photons from the progenitor
star and the cosmic ray escaping from the shock wave lastly ionize the cloud and generate thermal electrons,
it is possible to consider that both the hydrogen and carbon are ionized to contribute to a total electron density of 
$n_{e}\sim \frac{n_{\rm C^{+}}}{C/H}f(H_{2})+n_{\rm C^{+}}\sim$0.13/0.136 (H$^{+}$) + 0.13 (C$^{+}$)$\sim 1.09$~cm$^{-3}$.
With a filling factor of 1, the magnetic field strength in the thin layer of the cloud should be no less than 13~$\mu$G,
for the least intrinsic rotation measure of $-18$~rad~m$^{-2}$ required to have a depolarization effect in the model of~\citet{wr04}.

It has been observed that the stellar wind from the progenitor star of remnants has triggered star-formation in ambient molecular clouds
with young stellar objects (YSOs), such as IC~443~\citep{sfy14}. Although the CO cloud is not dense enough to form YSOs,
its shape resembles the fragmentation of CO clouds after the stellar wind of the progenitor star.

\section{Summary}
We have obtained new 21-cm continuum polarization cube data of the SNR HB~9 using the FAST radio telescope.
We decomposed the total-intensity image into components of multiple angular scales. It showed that the filamentary emission has a
spectrum with $\alpha=-$0.52, corresponding to the freshly accelerated relativistic electrons, and the
diffuse emission has a steeper spectrum with $\alpha=-$0.63, corresponding to the confined electrons that are no longer accelerated.

Depolarization associated with bright filaments in each shell has been observed in the FAST data at 1385~MHz.
The detected polarized emission might come from a thin layer in the outer envelope of the shells, with a Faraday depth of
$4-28$~rad~m$^{-2}$ from the Faraday rotation synthesis result.
We re-analyzed the rotation measures of HB~9 using the Effelsberg 2695-MHz and Urumqi 4800-MHz data. 
The RM is about 70~rad~m$^{-2}$ in the eastern and northern shell and $124$~rad~m$^{-2}$ in the inner and southern patches.
The regular magnetic field is estimated to be 5$-$8~$\mu$G over the remnant.

The northern shell shows depolarization at 2695~MHz with a depolarization factor of 0.6$-$0.7 relative to the 4800-MHz
polarization data. It can be explained by the beam depolarization with an $\delta RM$ of 36$-$43~rad~m$^{-2}$,
indicating an additional random magnetic field of 12~$\mu$G on the scale of 0.6~pc.
The shock might have entered the dense gas environment in the northern shell region and has driven turbulence to cause depolarization
at 2695~MHz.

\begin{acknowledgements}
We thank the journal referee for critical reading and valuable comments to improve the paper. 
This work is supported by the Key Program of National Natural Science Foundation of China (12433006), 
and the Guizhou Provincial Science and Technology Projects (No. QKHFQ[2023]003, No. QKHFQ[2024]001, No. QKHPTRC-ZDSYS[2023]003).
We acknowledge the help from FAST colleagues for using the server, and Dr. Li, J. T. for providing the MWISP CO data.
This research made use of the data from the Milky Way Imaging Scroll Painting (MWISP) project, which is a multi-line survey in 
12CO/13CO/C18O along the northern galactic plane with the PMO-13.7m telescope. We are grateful to all the members of the MWISP working group,
particularly the staff members at the PMO-13.7m telescope, for their long-term support. MWISP was sponsored by National Key R\&D Programme
of China with grants 2023YFA1608000 \& 2017YFA0402701 and by CAS Key Research Programme of Frontier Sciences with grant QYZDJ-SSW-SLH047.

\end{acknowledgements}

\bibliographystyle{aa}

\bibliography{G160.bib}

\end{document}